\documentclass[aps,twocolumn,showpacs]{revtex4}
\usepackage{latexsym}
\usepackage{graphics}
\usepackage{graphicx,amssymb,epsfig,epsf,amsmath}
\begin{document}
\title{Spectral analysis of molecular resonances in erbium isotopes:\\ Are they close to semi-Poisson?}
%\shorttitle{Spectral analysis of molecular resonances in Erbium isotopes} %Insert here a short version of the title if it exceeds 70 characters

\author{Kamalika~Roy$^{1}$, Barnali~Chakrabarti$^{1}$, N.~D.~Chavda$^{2}$, V.~K.~B.~Kota$^{3}$, M.~L.~Lekala$^{4}$,G.~J.~Rampho$^{4}$}

\affiliation{
$^{1}$Department of Physics, Presidency University,86/1 College Street, Kolkata 700 073, India.\\
$^{2}$Department of Applied Physics, Faculty of Technology and
Engineering, The Maharaja Sayajirao University of Baroda, Vadodara 390 001, India.\\
$^{3}$Physical Research Laboratory, Navarangpura, Ahmedabad 380009, India.\\
$^{4}$Physics Department, University of South Africa,  P.O.Box 392, Pretoria 0003, South Africa.
}

\begin{abstract}
We perform a thorough analysis of the spectral statistics of experimental molecular resonances, of bosonic erbium $^{166}$Er and $^{168}$Er isotopes, produced as a function of magnetic field($B$) by Frisch {\it{et al.}} [{\it{Nature {\bf{507}}, (2014) 475}}], utilizing some recently derived surmises which interpolate between Poisson and GOE and without unfolding. Supplementing this with an analysis using unfolded spectrum, it is shown that the resonances are close to semi-Poisson distribution. There is an earlier claim of missing resonances by Molina {\it{et al.}} [{\it{Phys. Rev. E {\bf{92}}, (2015) 042906}}]. These two interpretations can be tested by more precise measurements in future experiments.
\end{abstract}
\maketitle

\section{Introduction}
\label{intro}
There is no unique and precise definition of quantum chaos though it is closely related to spectral fluctuations and level correlations in a given quantum system. It has long been recognized from the study of atomic and nuclear spectroscopy that unambiguous determination of whether a quantum system is fully chaotic or intermediate between integrability and chaoticity is not possible in a straightforward manner as the data is usually limited for sufficient statistical significance. The ambiguity increases for strongly interacting and correlated many-body systems. The limiting case of quantum systems that are rotationally and time-reversely invariant and which have a fully chaotic classical analog, the level (also strength) fluctuations follow the predictions of the Gaussian orthogonal ensemble (GOE) of random matrix theory (RMT)\cite{BGS1984,Bohrev,Stoc,Casati1980}. This is known as Bohigas-Gionnoni-Schmit (BGS) conjecture.

There are enormous amount of numerical and experimental evidences  confirming the BGS conjecture and they are mainly from models for many-body quantum systems, nuclear energy levels, molecular spectra and so on \cite{Casati1980,Haake2010,SVZ1984,Zimm1986,Tanner2000,Sakhr2000,Br-81,VKBK2001,Gomez2011}. In many examples some deviations from RMT are observed. The deviations could be because the system may be intermediate to integrable and chaotic situation or the various levels/resonances in the observed spectrum may be carrying different symmetries or due to either of incomplete or imperfect spectra. In the incomplete spectra a fraction of levels may be unaccounted in the experiments, whereas in the imperfect spectra assignments of quantum numbers to the experimentally identified levels may be wrong. All these lead to different conclusions from the study of short-range and long-range spectral fluctuations with regard to intermediate statistics.

The trapped ultracold atoms and Bose-Einstein Condensation are the most thoroughly studied, highly complex, and correlated many-body systems where the atomic and molecular samples are reduced to temperatures of micro- to nano-Kelvin and the interaction between the particles can be manipulated easily. These resulted in the rich ultracold physics. Very recently the Innsbruck (Austria) group, Frisch {\it{et al.}} \cite{Frisch2014}, has studied collisions of trapped cold Erbium atoms, $^{166}$Er and $^{168}$Er isotopes, as a function of the magnetic field ($B$). In the experiment of ultracold collision of gas-phase Erbium atoms, many Feno-Feshbach resonances were exhibited. It is the first and the only one experimental demonstration of random spectra at ultracold temperature; see \cite{DYS} for other attempts. To calculate the spectrum of Fano--Feshbach resonances in \cite{Frisch2014}, a first-principles coupled-channel (CC)model for Er--Er
scattering is constructed which uses the atomic basis set and Hamiltonian
(Methods) that includes the radial kinetic and rotational energy operators,
the Zeeman interaction and anisotropic B--O potentials.
For small interatomic separations $R$, the B--O potentials are calculated
using the {\it ab initio} relativistic, multi-reference configuration –interaction
method. At intermediate to large values of $R$, the B--O potentials are
expressed as a sum of multipolar interaction terms. The van der Waals
dispersion interaction potentials are determined from experimental
data on atomic transition frequencies and oscillator strengths. Here, the dispersion potentials have both isotropic
and anisotropic contributions. Frisch {\it{et al.}} showed that the CC method together with a random quantum defect theory describes the mean density of the resonances quite well. As they state, 'given the complexity
of the scattering, the analysis of ultracold collision data
can not and should not aim anymore at the assignment
of individual resonances'. Following this, they made a statistical analysis of the resonances using the tools of RMT and found intermediate statistics between Poisson and Wigner-Dyson (WD) or GOE. No definite conclusion has been made and as the statistical analysis are close to WD statistics, it was concluded that the dynamics of collisions and formation of Er-Er molecules is very complex and presents characteristics of quantum chaos \cite{Frisch2014}. Later, a more detailed analysis of the same spectra has been done by Petit and Molina \cite{Molina2015} from the view point of missing resonances using nearest neighbour spacing distribution (NNSD) modeled by the so called Brody distribution, power spectrum and  statistics of resonance widths. They also found intermediate statistics between Poisson and WD and concluded that the disagreement with RMT is due to the possibility of missing resonances (about 20\%) in the spectrum. This is also to be noted that some earlier works in the context of chaos in ultracold atomic systems have been reported using the standard Bose-Hubbard model \cite{Buch2003,Tom2007,Tom2009,Buon2008,CAP2013,DF2016,ARK2004,Dub2016}. Here, mainly different set-ups are analyzed and many related questions on long-range correlations and symmetries are addressed. However, none of them are related with the molecular resonance spectra from ultracold collisions.

In the present letter, we extend the analysis of spectral statistics in experimental molecular resonances of bosonic Erbium $^{166}$Er and $^{168}$Er with the aim of identifying that the intermediate statistics is close to the so called semi-Poisson \cite{semi-pois-1}. To this end, we have employed the recently introduced measure of ratio of nearest spacings, by Oganesyan and Huse~\cite{OH2007},  that do not require unfolding. This analysis is supplemented with more traditional analysis with unfolding.
The resonance data for isotopes $^{166}$Er and $^{168}$Er are presented in Extended Data Tables 1 and 2 at the end of ref.~\cite{Frisch2014}. These tables report the Fano-Feshbach resonance positions $B$ and width $\Delta$. We have directly utilized these data sets for the statistical analysis ignoring the resonance widths and no further extraction process is done.
Now, we will turn to the analysis using ratio of nearest spacings.

\section{Statistical analysis using $P(r)$ distribution}
\label{sec2}
Generally, NNSD $P(s)ds$ giving degree of level repulsion and Dyson-Mehta $\Delta_3$-statistics or the power spectrum with $1/k^\alpha$ noise giving long-range spectral rigidity, are important measures in the study of level
statistics. In constructing NNSD or the power spectrum (similarly $\Delta_3$ statistic) for a given set of energy levels, the spectra have to be unfolded to remove the secular variation
in the density of eigenvalues \cite{Haake2010,Br-81}. The final outcome of NNSD may vary with the procedure of unfolding. We do not know in general {\it{a priori}} the proper unfolding function for most of the realistic systems and therefore generally it is approximated by higher order polynomials. On the other hand, Oganesyan and Huse~\cite{OH2007} considered the distribution of the ratio of consecutive level spacings of the energy levels ($P(r)$) which requires no
unfolding as it is independent of the form of the density of the energy levels. The $P(r)$ distribution gives more transparent comparison with experimental results than traditional level-spacing distribution. This measure is specifically fruitful for many-body systems as the theory of eigenvalue density for these systems is usually not available~\cite{sudpra14}. In recent past, this measure was used in analyzing many-body localization~\cite{OH2007,OPH2009,pal,iyer} and also in quantifying the deviation from integrability on finite-size lattices~\cite{kol,col}. Using $P(r)$, it has been concluded that the embedded random matrix ensembles for many body systems, generated by random interactions in presence of a mean-field, follow GOE for strong enough two-body interaction~\cite{CK}. Very recently,  in \cite{Dietz2016} $P(r)$ distributions of rectangular and Africa-shaped superconducting microwave resonators containing circular scatterers on a triangular grid, so-called Dirac billiards (DBs) are studied. It is shown that for statistics near the van Hove singularities the ratio distributions provide suitable measures. In next section, we analyze the spectra using the $P(r)$ distributions.

\subsection{Results for $P(r)$, $P(\tilde{r})$, $\langle r\rangle$ and $\langle \tilde{r}\rangle$ and comparison with semi-Poisson}

For an ordered set of energy levels $E_n$, the nearest level spacing is $S_n=E_{n+1}-E_{n}$ and the probability distribution of the ratios $r_n=\frac{S_n}{S_{n-1}}$ is $P(r)$ subject to normalization $\int P(r)dr=1$. In the integrable domain, NNSD follows Poisson distribution and in this region the $P(r)$ is given by~\cite{OH2007,Atas2013,prkota}
\begin{equation}
 P_{P}(r)=\frac{1}{(1+r)^2}\;.
\end{equation}
Similarly, in the chaotic region the distribution follows Wigner-Dyson (WD) or GOE statistics~\cite{Atas2013,prkota} and it is
\begin{equation}
P_{GOE}(r)=\frac{27}{8}\frac{r+r^2}{(1+r+r^2)^{5/2}}\;.
\end{equation}
In addition, the average value of $r$, i.e., $\langle r \rangle$ is 1.75 for GOE and is $\infty$ for Poisson. It is also possible to consider $\tilde{r} = \frac{min(S_n, S_{n-1})}{max(S_n,S_{n-1})} = min(r_n, 1/r_n)$. As pointed out in \cite{Atas2013}, it is possible to write down, the distribution of $\tilde{r}$, $P(\tilde{r})$, for a given $P(r)$. The average value of $\tilde{r}$ i.e. $\langle \tilde{r} \rangle$ is 0.536 for GOE and 0.386 for Poisson~\cite{Atas2013}.

\begin{figure}
  \begin{center}
\includegraphics[width=\linewidth]{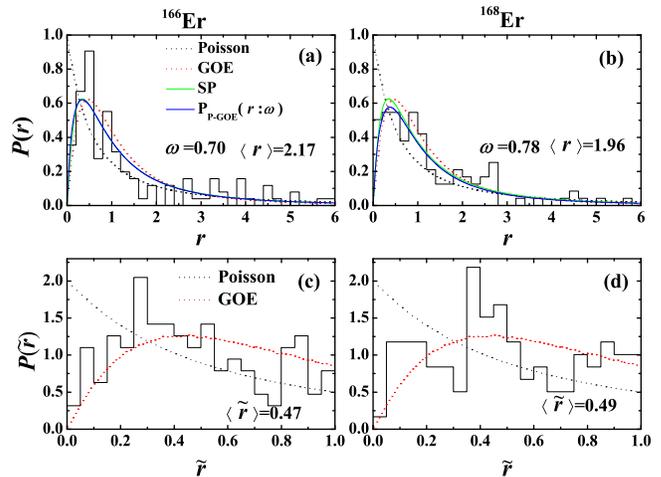}
  \end{center}
\caption{Plot of $P(r)$ distributions for (a) $^{166}$Er and (b) $^{168}$Er; $P(\tilde{r})$ distributions for (c) $^{166}$Er and (d) $^{168}$Er.}
\label{pr}
\end{figure}

In Fig. \ref{pr}, the results for $P(r)$  and $P(\tilde{r})$ distributions for Erbium are shown. Here, also for comparison, results from $P_{P}(r)$ and $P_{GOE}(r)$ are shown in Fig. \ref{pr} (a) and (b). More importantly, they are also compared with the $P(r)$ for generalized semi-poisson distribution (GSP) as given in \cite{Atas2013a},
\begin{equation}
P_{\nu}(r)=\frac{\Gamma(2\nu+2)\Gamma^2(\nu+2)}{(\nu+1)^2 \Gamma^4(\nu+1)}\frac{r^{\nu}}{r^{2\nu+2}}\;.
\end{equation}
Here $\nu$ is a parameter with $\nu=1$ giving semi-Poisson (SP).
The NNSD $P(s)$ for the semi-Poisson is $P(s)=4s\exp(-2s)$ and the NNSD for the generalized semi-Poisson is given ahead in Eq. (\ref{sp}).
It is well established that the SP distribution is appropriate for pseudo-integrable systems \cite{semi-pois-1,semi-pois-2,semi-pois-3} and also for the level statistics (critical statistics) at metal-insulator transition (MIT) point. In addition to $P(r)$, also shown are results for $P(\tilde{r})$ vs$\langle \tilde{r} \rangle$ in Fig. \ref{pr} (c) and (d).

The results of $P(r)$ and $P(\tilde{r})$ distributions are seen to be intermediate between Poisson and GOE and close to SP. We have also calculated $\langle r \rangle$ and $\langle \tilde{r} \rangle$. For $^{166}$Er, $\langle r \rangle = 2.17$ and $\langle \tilde{r} \rangle = 0.47$, whereas for $^{168}$Er, $\langle r \rangle = 1.96$ and $\langle \tilde{r} \rangle = 0.49$. For SP, the average values are $\langle r \rangle = 2$ and $\langle \tilde{r} \rangle = 0.5$. These are close to the values calculated for both $^{166}$Er and $^{168}$Er.

In order to further quantify the departures from GOE, we use the Poisson to GOE interpolation form given in \cite{CK}
\begin{equation}
 P_{P-GOE}(r:\omega) = \frac{1}{Z_\omega}\;\frac{(r + r^2)^\omega}{ \left(1+ (2-\omega)r +r^2\right)^{1+\frac{3}{2}\omega}}\;.
\end{equation}
Here, $\omega=0$ gives Poisson and $\omega=1$ gives GOE correctly. The parameter $Z_\omega$ is obtained using the condition
$\int^{\infty}_0 P(r) dr =1$. Best fit to data gives  $\omega = 0.70$ for $^{166}$Er and $\omega = 0.78$ for $^{168}$Er. It is important to note that for SP, $\omega \sim 0.7$. This is also in good agreement with our previous results of $P(r)$  and $P(\tilde{r})$, indicating clearly that the statistical features of the resonance data for  $^{166}$Er and  $^{168}$Er are intermediate between Poisson and GOE and close to SP.

\begin{figure}
  \begin{center}
\includegraphics[width=\linewidth]{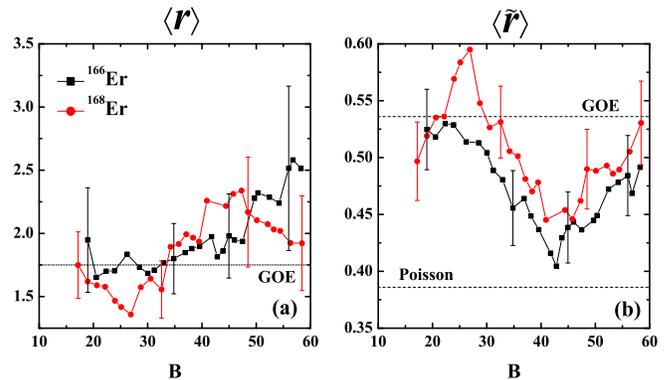}
  \end{center}
\caption{Average values of (a) $r$ and (b) $\tilde{r}$ vs. magnetic field obtained using a moving window of 60 levels for $^{166}$Er (black squares)
and $^{168}$Er (red circles). The dashed horizontal lines indicate the limiting values corresponding to Poisson and GOE estimates. The widths of distribution of $r$ and $\tilde{r}$ are shown by vertical bars for only few points in the figure for clarity.}
\label{avg}
\end{figure}

To further investigate the chaoticity of the two systems, computed are $\langle r \rangle$ and $\langle \tilde{r} \rangle$ for the spectra as a function of the external magnetic field. Following \cite{Molina2015}, we divide the spectra in windows of $60$ levels (resonances) around a particular magnetic field and computed the corresponding $\langle r \rangle$ and $\langle \tilde{r} \rangle$. The results are shown in Fig. \ref{avg}. Corresponding error-bars in some points are mentioned in the figures for clarity. It is to be noted that the average value of $r$, i.e. $\langle r \rangle$ changes from 1.75 to $\infty$ that  corresponds to transition from GOE to Poisson. Whereas the average value of $\tilde{r}$, i.e., $\langle \tilde{r} \rangle$ changes from 0.536 to 0.386 for the transition from GOE to Poisson. From Fig. \ref{avg} (a), we observe that the measures $\langle r \rangle$ and $\langle \tilde{r} \rangle$ give the transition near $B\approx30$ G. However, deviations exist both at smaller and higher values of $B$. Also for $B>40$,  $\langle r \rangle$ and $\langle \tilde{r} \rangle$ values are close to that of the SP distribution.

%..............fig.~1

%..............fig.~3

\section{Statistical analysis using unfolded spectra}
\label{sec3}
Statistical analysis of the experimental data is carried on further using the unfolding procedure. Here the eigen-energies $E_i$ of a system are transformed to new eigen-energies $\epsilon_i$ keeping the density of levels locally constant with unit mean~\cite{unfo3}. The integrated level density $N(E)$ has two parts, a smooth part [$\overline{N}(E)$] and a fluctuating part [$\tilde{N}(E)$]. To compare the fluctuation properties of different systems or different parts of the same system, the smooth part is removed using a unfolding procedure. Unfolding removes the variation in the density of the levels in different parts of the spectrum.
In the present analysis our focus is mainly on the spectra for magnetic field $B \geq$ 30 G because below 30 G of magnetic field there is a steady increase of chaoticity \cite{Frisch2014}. In the unfolding procedure, we have approximated the level density by a fifth order polynomial for the full spectrum and a linear fit to data for $B \ge 30$G for both isotopes as it was done in ~\cite{Molina2015}.

\subsection{Results for NNSD}

\begin{figure}
  \begin{center}
\includegraphics[width=\linewidth]{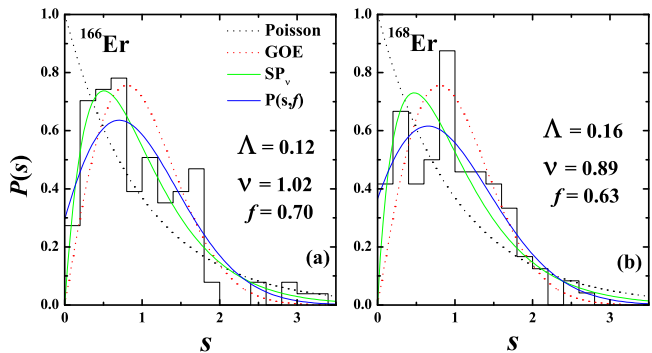}
  \end{center}
\caption{Plot of NNSD, $P(s)$ for (a) $^{166}$Er and (b) $^{168}$Er. They are compared to the Poisson (black doted line) and the WD
(red doted line) distribution. The solid curves in blue and green were determined using distributions given by Eqs. (\ref{psf}) and (\ref{sp}) respectively. The best fit parameter values are also given for both the distributions.}
\label{nnsd}
\end{figure}

Using the unfolded eigenvalues $\epsilon_{i}$, the nearest neighbor spacing is calculated as $s_i = \epsilon_{i+1} - \epsilon_{i}$, $i=1,2,...,n$. The probability distribution $P(s)$ of these spacings is shown for both $^{166}$Er and $^{168}$Er in Fig.~\ref{nnsd} (a) and Fig.~\ref{nnsd} (b) respectively. It is well known that when the system is in integrable domain, the form of NNSD
is close to the Poisson distribution \cite{Bohrev,Stoc}, i.e.
\begin{equation}
P_P(s) = \exp(-s)\;.
\label{p}
\end{equation}
while in the chaotic
domain or for GOE, the form of NNSD is the Wigner surmise or the Wigner-Dyson (or GOE) ~\cite{Bohrev,Stoc}
\begin{equation}
 P_{WD}(s) = (\pi s/2)\; \exp (-\pi s^2/4)\;.
 \label{wd}
\end{equation}
Going further, the NNSD form for the GSP distribution is given by \cite{semi-pois-2},
\begin{equation}
 P_{\nu}(s) = \frac{(\nu+1)^{\nu+1}}{\Gamma(\nu+1)} s^{\nu}\; \exp (-(\nu+1) s)\;.
 \label{sp}
\end{equation}
Here, $\nu$ is a parameter. More importantly, $\nu=1$ gives, as mentioned before, the NNSD for the semi-Poisson
\begin{equation}
P_{\nu=1}(s)=4s \exp(-2s)\;.
\end{equation}
Results for the best fit GSP distribution are shown in Fig. \ref{nnsd} and also the NNSD form the unfolded spectra are compared with Poisson and GOE forms. Value of the parameter $\nu$ obtained is $1.02$ for $^{166}$Er and $0.89$ for $^{168}$Er. From these results, it is clear that the NNSD for both $^{166}$Er and $^{168}$Er is very close to the semi-Poisson $P_1(s)$ distribution and it is intermediate to Poisson and Wigner-Dyson distributions. Note that the level repulsion at smaller values of $s$ ($s \ll 1$), is $\propto s$ for both $P_{WD}(s)$ and $P_1(s)$ and asymptotic decay of $P_1(s)$ is exponential decay.

The deviations from the WD form for NNSD is usually discussed in terms of the Brody \cite{Br-81,iz} and Berry and Robnik distributions\cite{Berry1984}. The former enjoys considerable popularity but its interpolating
parameter lacks to have a deeper physical
meaning. However the Berry–Robnik parameter $\rho$, has
been shown to be the fraction of the regular phase
space domain. Here, we have chosen yet simple criterion given in ref.\cite{Ko-99}. In  \cite{Ko-99} using an appropriate $2\times2$ random matrix model, transition curve for the variance $\sigma^2(0)$ of the NNSD is constructed for the Poisson to GOE transition
in terms of the parameter $\Lambda$. The Poisson value $\sigma^2(0)=1$ and for GOE value for $\sigma^2(0)= 0.27$ and then the $2\times2$ random matrix
model gives $\Lambda=0$ for Poisson and $\Lambda \sim 1$  for GOE. There is an
onset of GOE fluctuations at $\Lambda \sim 0.3$ and the formula
for $\sigma^2(0:\Lambda)$ \cite{Ko-99} gives $\sigma^2(0)=0.37$ for $\Lambda = 0.3$. In figure \ref{nnsd}, the  histograms show results for NNSD  obtained for Er data and these are compared with WD and Poisson distributions. The value of $\Lambda$ is found to be 0.12 for $^{166}$Er and 0.16 for $^{168}$Er and these values are much smaller than the critical value 0.3. However, $\sigma^2(0)$ for $P_1(s)$ is $0.5$ and the corresponding value of $\Lambda=0.12$ is in good agreement with the $\Lambda$ value obtained for $^{166}$Er and not far from the value obtained for $^{168}$Er.

Further, we have investigated the chaoticity of the system as a function of the external magnetic field using unfolded energy spectra. Here, dividing the spectra in windows of $60$ levels around a particular magnetic field and computed the variance $\sigma^2(0)$ of NNSD and the results are shown in Fig. \ref{var_nnsd}. It is clear from the figure that for the weak magnetic field, $\sigma^2(0)$ values of the spectra are above critical value 0.37. And as the magnetic filed increases, the spectral fluctuations reach close to GOE for $B\lesssim30$G. Further increase in magnetic field leads to the transition from GOE to Poisson near $B\approx30$G. These results are consistent with that of obtained using the ratio of spacings $r$ analysis carried out in the previous section.

We have also carried out an analysis of the spectra using the
hypothesis that the observed spectrum is a superposition of resonances with different symmetries.
The NNSD in this situation, with some approximations,is  given in terms of a single parameter $f$ \cite{Abul1996,Dietz2017},
\begin{equation}
 P(s,f) = \left\{ 1-f + \frac{\pi}{2} Q(f)s \right\}\; \exp \left\{-(1-f)s+\frac{\pi}{4} Q(f)s^2 \right\}\;.
 \label{psf}
\end{equation}
with $Q(f)= 0.7f + 0.3 f^2$. Note that, for $f=0$, the distribution approaches the Poisson distribution, which corresponds to a spectrum composed of a large number of sub-spectra of GOE type but each consisting of a few number of levels. Whereas, $f=1$ gives the NNSD of the GOE. Therefore, $f$ is referred to as the chaoticity parameter. The blue curves in fig.\ref{nnsd} are obtained using Eq. (\ref{psf}). The chaoticity parameter $f$ values found are $0.70$ for $^{166}$Er and $0.63$ for $^{168}$Er. These results also show deviation from GOE. The best fit of $P_1(s)$ to $P(s,f)$ gives $f=0.72$ which is also close to the values of $f$ obtained in both the examples.

\subsection{$\Delta^2$ test}

\begin{figure}
  \begin{center}
\includegraphics[width=0.7\linewidth]{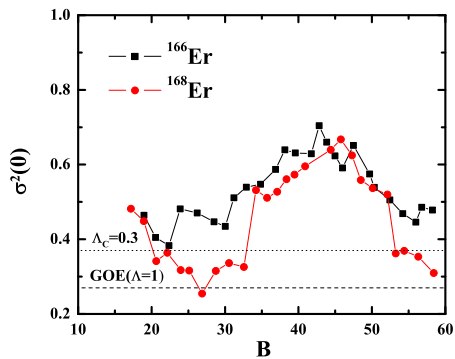}
  \end{center}
\caption{Variance of NNSD, $\sigma^2(0)$ plotted as a function of magnetic field obtained using a moving window of 60 levels for $^{166}$Er (black squares) and $^{168}$Er (red circles). For Poisson distribution $\sigma^2(0)=1$ while for GOE $\sigma^2(0)=0.27$(shown by dash line). Also the critical value for Poisson to GOE transition $\sigma^2(0)=0.37$ is shown by dotted line.}
\label{var_nnsd}
\end{figure}

\begin{figure}
  \begin{center}
\includegraphics[width=\linewidth]{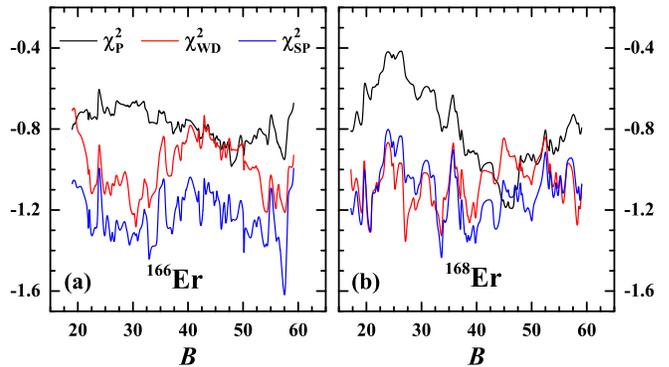}
  \end{center}
\caption{$\Delta^2$ statistical tests for (a) $^{166}$Er and (b) $^{168}$Er.}
\label{chi2}
\end{figure}

Further, we use a $\Delta^2$ test to measure the distance from the obtained numerical data to the theoretical predictions from the Eqs.(\ref{p}), (\ref{wd}) and (\ref{sp}) as a function of external magnetic field as done in previous section using ratio of consecutive levels. The $\Delta^2$ is defined as,
\begin{equation}
\Delta^2_{\alpha}=\log_{10} \left\{ \int_{0}^{\infty} ds\; ( P_{\alpha}(s) -  P(s) )^2 \right\}\;.
\label{del2}
\end{equation}
Where $\alpha$ stands for Poisson or Wigner-Dyson or semi-Poisson $P_1(s)$ respectively. Here, we have divided the spectra in windows of 60 levels (using unfolded levels) around a particular magnetic field and then $\Delta^2$ is computed via $P(s)$ distribution. Lower value of $\Delta^2$ measure indicates the closeness of $P(s)$ with the theoretical prediction. The results for $\Delta^2$ test are shown in Fig. \ref{chi2} for both $^{166}$Er and $^{168}$Er. For $^{166}$Er, the lowest values for $\Delta^2_{SP}$ for entire range of magnetic field (Fig. \ref{chi2}(a)). For $^{168}$Er, $\Delta^2$ values for SP and WD are close upto $B<25$ G and beyond this, $\Delta^2$ values are minimum for SP. Hence, These results conform the closeness of the NNSD with the SP distribution compared to Poisson and Wigner distribution for both $^{166}$Er and $^{168}$Er.
%\hspace*{1cm}
Before going further, with regard to the results presented in Figs.~\ref{avg}, \ref{var_nnsd} and \ref{chi2}, it is important to mention that for each $B$ value, we are using $60$ resonances (for having enough statistics), i.e. roughly $20$G wide region as there are approximately $3$ resonances per Gauss in the data. The range for $B$ in the figures is $60$G and therefore for each $B$ value used is roughly a third of the whole range presented in these figures. With this, one can argue that there are only three uncorrelated data points in these figures. In this respect the result that there is a transition at $B \sim 30$G need to be used with caution.
%..............fig.~4

\subsection{Power spectrum analysis}

Long-range correlations among energy levels are investigated using the analogy between energy spectrum and discrete time series~\cite{jmg,arel,arel1,efa}. The energy spectrum can be considered as a discrete signal and then the fluctuations in the excitation energies generate a discrete time series. The $\delta_n$-statistics of consecutive levels,
\begin{equation}
\delta_{n} = \sum_{i=1}^{n}(s_{i}-<s>) = \epsilon_{n+1}-\epsilon_{1}
-n \;,
\end{equation}
is similar to the time-series and $n$ represents the discrete time.
Now, the power spectrum is defined as the square modulus of the Fourier transform of $\delta_n$ giving,
\begin{equation}
P_{k}^{\delta} = |{ \frac{1}{\sqrt{M}} \sum_{n} \delta_{n} exp(-\frac{2\pi ikn}{M})}|^{2}\;.
\label{power}
\end{equation}
Here, $k=1,2,...,n$ and $M$ is the size of the series ~\cite{jmg}. Therefore, the statistical behavior of the level fluctuations can be investigated by using $ \langle P_{k}^{\delta}\rangle$ statistics which measures  both short and long range correlations.

\begin{figure}
\begin{center}
\includegraphics[width=0.7\linewidth]{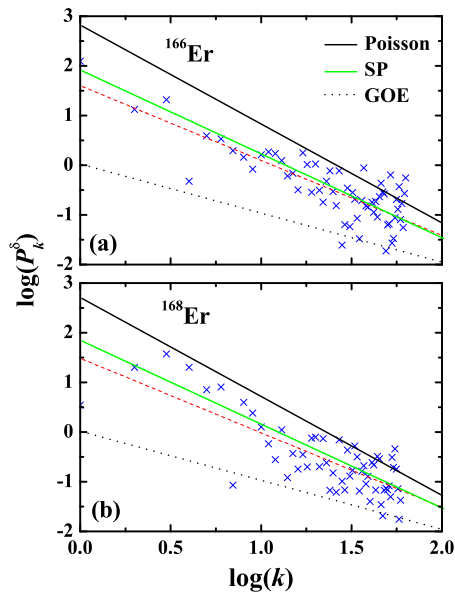}
\end{center}
\caption{(color online): Power spectrum of  $\delta_n$ for (a) $^{166}$Er and (b) $^{168}$Er. The red dashed lines are the best fit $\langle P_{k}^{\delta}\rangle \propto \frac{1}{k^{\alpha}}$. The black line and  the green line superimposed in each graph are the predictions of Poisson and semi-Poisson respectively while the dotted lines in the each graph are predictions of GOE.}
\label{noise}
\end{figure}

Using the $^{166}$Er and $^{168}$Er data, $\delta_n$ is calculated as a function of $n$. Then, using Eq.~(\ref{power}), the power-spectrum is generated and the results are shown in Fig.~\ref{noise}. Power-spectrum shows $\frac{1}{f}$ noise for pure chaos and $\frac{1}{f^{2}}$ noise for pure integrable systems~\cite{jmg,arel,arel1,arel2}, where $f=\frac{2\pi k}{M}$. As seen from Fig. \ref{noise}, Erbium systems exhibit intermediate nature between integrability and chaos. Following our earlier study of $^{87}$Rb~\cite{pre2012}, we have fitted the power spectrum to $\frac{1}{k^{\alpha}}$ curve and extracted the constant $\alpha$; see Figs.~\ref{noise}(a) and \ref{noise}(b). Here $\alpha$ measures the degree of chaoticity and the fits give $\alpha=1.51 \pm 0.15$ for $^{166}$Er and $\alpha=1.50\pm 0.15$ for $^{168}$Er, which are intermediate to Poisson and GOE. The expected $\alpha$ value for SP is $\sim 1.7$ \cite{AMG2006} which is close to the $\alpha$ values obtained for both the examples compared to GOE predictions. Thus, the power spectrum also shows SP structure.

%..............fig.~6

\section{Conclusions}
\label{sec4}
In this paper presented is an exhaustive study of the spectral statistics of experimental molecular resonances in bosonic $^{166}$Er and $^{168}$Er atoms. The resonances are produced as a function of the magnetic field by Frisch {\it{et al.}} \cite{Frisch2014}. In the analysis we have used the spectrum without unfolding and used the new measure of ratio of nearest spacings. It is clearly seen that the resonances follow semi-Poisson. For re-confirmation of this results we have also used the unfolded spectrum and analyzed NNSD and power spectrum. We have utilized several interpolating formulas for Poisson to GOE and $\Delta^2$ giving mean-square difference between measured NNSD and P, WD and SP distributions. All these clearly confirm that the Erbium resonances follow Semi--Poisson distribution. This conclusion is different from Molina {\it et al.} \cite{Molina2015} claim that the observed deviations from GOE are due to missing resonances. Increasing magnetic-field resolution and producing higher-quality data will clearly confirm if there are missing resonances. Also a negative result of such an experiment will prove the result of the present paper that the
resonances are of semi-poisson character. Also experiments with other molecules will be useful; see for example \cite{DYS} for attempts in this direction.

\acknowledgments
KR acknowledges Council of Scientific and Industrial Research (CSIR), India, for her senior research fellowship [grant~no.~08/155(0049)/2015-EMR-I]. BC thanks Department of Science and Technology, both of Government of India [fund~no.~SR/S2/CMP/0126/2012] and Government of West Bengal [memo~no.~1211(Sanc.)/ST/P/S\&T/4G-1/2012] for financial support. NDC acknowledges support from Department of Science and Technology, Government of India [project~No:~EMR/2016/001327].

\end{document}